\documentclass[%
reprint,
amsmath,amssymb,
aps,article
]{revtex4-2}
\usepackage{CJKutf8}
\usepackage{graphicx}
\usepackage{dcolumn}
\usepackage{bm}
\usepackage{color} 
\usepackage {braket}
\usepackage{float}
\usepackage{mathrsfs}
\usepackage{amssymb}
\usepackage{bbm}
\usepackage{amsmath}
\usepackage[colorlinks, citecolor=blue]{hyperref}

\begin{document}
	
	\preprint{APS/123-QED}
	
	\title{The Security Analysis of Continuous-Variable Quantum Key Distribution under Limited Eavesdropping with Practical Fiber}
	
	\author{Sheng Liu\textsuperscript{1}, Lu Fan\textsuperscript{2}, Zhengyu Li\textsuperscript{3}, Qiang Zhou\textsuperscript{4}, Yunbo Li\textsuperscript{1}, Dong Wang\textsuperscript{1}, Dechao Zhang\textsuperscript{1}}\author{Yichen Zhang\textsuperscript{2}}\email{zhangyc@bupt.edu.cn}\author{Han Li\textsuperscript{1}}
	\address{%
		\textsuperscript{1}Department of Fundamental Network Technology, China Mobile Research Institute, Beijing, China\\
		\textsuperscript{2}State Key Laboratory of Information Photonics and Optical Communications, School of Electronic Engineering,
		Beijing University of Posts and Telecommunications, Beijing 100876, China\\
		\textsuperscript{3}Central Research Institute, 2012 Labs, Huawei Technologies Co., Ltd, Shenzhen 518129, Guangdong, China\\
		\textsuperscript{4}Institute of Fundamental and Frontier Sciences, University of Electronic Science and Technology of China, Chengdu 611731, China
	}%

	
	\date{\today}
	
	\begin{abstract}
		
		Research on optimal eavesdropping models under practical conditions will help to evaluate realistic risk when employing quantum key distribution system for secure information transmission. Intuitively, fiber loss will lead to the optical energy leaking to the environment, rather than harvested by the eavesdropper, which also limits the eavesdropping ability while improving the quantum key distribution system performance in practical use. However, defining the optimal eavesdropping model in the presence of lossy fiber is difficult because the channel is beyond the control of legitimate partners and the leaked signal is undetectable.
		Here we investigate how the fiber loss influences the eavesdropping ability based on a teleportation-based collective attack model which requires two distant stations and a shared entanglement source. We find that if the distributed entanglement is limited due to the practical loss, the optimal attack occurs when the two teleportation stations are merged to one and placed close to the transmitter site, which performs similar to the entangling-cloning attack but with a reduced wiretapping ratio.
		Assuming Eve uses the best available hollow-core fiber, the secret key rate in the practical environment can be $20\%-40\%$ higher than that under ideal eavesdropping.
		While if the entanglement distillation technology is mature enough to provide high quality of distributed entanglement, the two teleportation stations should be distantly separated for better eavesdropping performance, where the eavesdropping can even approach the optimal collective attack. 
		Under the current level of entanglement purification technology, the unavoidable fiber loss can still greatly limit the eavesdropping ability as well as enhance the secret key rate and transmission distance of the realistic system, which promotes the development of quantum key distribution systems in practical application scenarios.	
		
	\end{abstract}
	
	\maketitle
	
	
	\section{\label{sec:level1_Into}Introduction}
	Quantum key distribution (QKD) \cite{scarani2009security, pirandola2020advances, xu2020secure} generates secure keys between legitimate partners, which is one of the most promising quantum communication protocols to reach the maturity for commercialization. Continuous-variable (CV) QKD \cite{grosshans2002continuous, weedbrook2004quantum, zhang2023continuous} has received great attentions, due to its better compatibility with classical optical communication devices \cite{guo2021toward} and potential high key generation rates in metropolitan
	areas \cite{jouguet2013experimental, jouguet2012field, huang2016field, zhang2019continuous, zhang2020long, wang2022sub, jain2022practical, pi2023sub, brunner2023demonstration, roumestan2022experimental, hajomer2023continuous}.
	The theoretical security poof of CV-QKD protocols with Gaussian states was firstly performed in the asymptotic limit \cite{garcia2006unconditional, navascues2006optimality}, and then extended to the finite-size situation \cite{renner2009finetti, leverrier2013security, pirandola2021composable, pirandola2021limits}.
	
	 QKD technology has attracted telecommunication operators' interests for a long time, and recently operators start to deploy QKD networks aimming at providing quantum key to customers as a service. Therefore, beside the theoretical security, it is also important to analyze the practical security of QKD systems in real application. This usually results in two categories of discussions. 
	 
	 The first is about side channels of QKD systems and the countermeasures. The main side-channel attacks against CV-QKD are targeting at detectors, such as local oscillator attack \cite{jouguet2013preventing, PhysRevA.88.022339, fan2023quantum}, wavelength attack \cite{huang2013quantum, ma2013wavelength}, blind attack \cite{qin2018homodyne}, polarization attack \cite{zhao2018polarization} and reference pulse attack \cite{shao2021phase, shao2022phase}.
	Defending methods include adding system monitors such as local oscillator monitoring, as well as the CV measurement-device-independent (CV-MDI) system \cite{li2014continuous, pirandola2015high, tian2022experimental} proposed immune to any detection attacks.
	The second is about limited eavesdropping under more realistic technology assumptions, such as individual attack \cite{lodewyck2007tight, sudjana2007tight}, the eavesdropping without quantum memory \cite{bechmann2006eavesdropping}, and restricted wiretapping attack \cite{pan2020secret}.
	
	These discussions about practical security are helpful for understanding the real risks from the open channel, that one may face for a practically deployed QKD system. This may also be potentially linked to Quality of service (QoS) classification of QKD service in the future. Higher QoS level requires security under less constraints on eavesdropper's abilities, thus usually higher service pricing. 
	
	In this paper, we investigate the limitation and impact of fiber intrinsic loss introduced to the practical eavesdropping ability.
	In the theoretical security analysis, it is assumed that Eve can fully control the fiber channel by replacing it with lossless fiber or harvesting all the lost energy, whereas is highly infeasible in practice. The generic eavesdropping models used for individual attack \cite{lodewyck2007tight, sudjana2007tight} and collective attack \cite{garcia2006unconditional, navascues2006optimality} are not enough for the discussion of limited eavesdropping with practical fibers. Recently, a teleportation-based collective attack \cite{tserkis2020teleportation} was proposed with two distant stations, in which Eve does not require lossless fibers, but relies on distributed entanglement. The attack strength varies between individual attack and optimal collective attack, according to the entanglement distributed between Eve's two stations.
	
	Based on this teleportation-based attack model, we further investigate when the fiber loss limits the entanglement distribution and how the eavesdropping ability is. Thus it provides a baseline for the discussion of limited eavesdropping with non-zero fiber loss. 
	We find that, when the entanglement distilled between Eve's two stations is limited due to the fiber loss, the optimal attack occurs when Eve's two stations merged to one and placed close to transmitter site, which performs similar to the entangling cloning attack \cite{10.5555/2011564.2011570} but with a reduced wiretapping ratio.
	With such the optimal attack in practical environment, the secret key rate will be much higher compared to the ideal eavesdropping situation.
	While if the entanglement distributed is large enough by applying the probabilistic noiseless linear amplifier, Eve's two stations should be separated distantly, with the performance still approaching to the optimal collective attack.
	
	
	The paper is structured as following.
	In Section II, we introduce the optical teleportation based attack model, and briefly analyze the limitations, including the fiber loss.
	In Section III, We analyze the limited eavesdropping model with practical fiber loss, in which we apply the noiseless linear amplifier to relax the limitation. 
	We finally have a discussion in Section IV.
	
	\section{\label{sec:level1_model}General eavesdropping model with optical teleportation}

	
	Traditional CV teleportation requires instant individual measurements for each transmission of quantum signal. While in all-optical CV teleportation, no instant measurements are required. Thus, the teleportation-based attack \cite{tserkis2020teleportation} with all optical teleportation architecture can reach optimal collective attack. For convenience, we briefly introduce the teleportation-based eavesdropping model and its limitations when practically used.
	
	\subsection{Eavesdropping Model Description}
	\begin{figure}[t]
		\includegraphics[width=8cm]{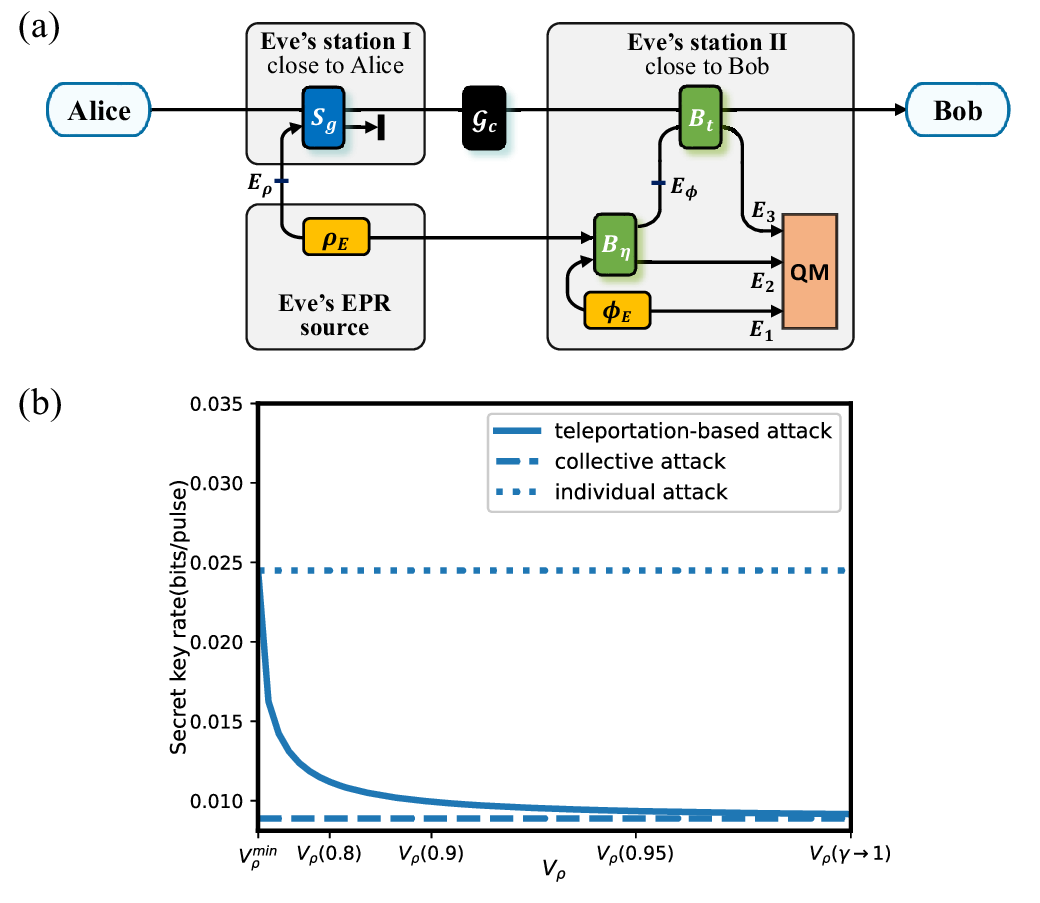}
		\caption{\label{fig1}(a) The teleportation-based collective attack. Eve prepares EPR state $\rho_E$ shared by station I and II. In station I, $S_g$ is the two mode squeezing operation with gain $g\textgreater1$.
			The output signal mode goes through a fiber channel to station II. 
			In station II, Eve prepares the second EPR state $\phi_E$ to simulate channel noise. The beamsplitters $B_{\eta}$ and $B_t$ have transmittances as $\eta$ and $t$, respectively. (b) Simulation of the influence from the variance $V_{\rho}$ of $\rho_E$ on the secret key rate. The dotted line represents the individual attack, the dashed line describes the collective attack, and the solid line describes the teleportation-based attack. The simulated transmission channel $\mathcal{G}_{equ}$ has 50 km long with attenuation coefficient 0.275 dB/km, and excess noise is $\epsilon=0.04$. Other simulation parameters are, Alice's modulation variance $V_A=4$, and reconciliation efficiency $\beta=0.96$.
		}
	\end{figure}

	When discussing a specific eavesdropping model, it should be able to simulate the same channel parameters as what the transmitter and receiver can estimate. In CV-QKD, the most commonly used security analysis method is based on Holevo bound \cite{garcia2006unconditional,holevo1973information} derived from Gaussian extramlity theorem \cite{wolf2006extremality}, in which the channel parameters are mainly equivalent transmittance $T_{equ}$ and equivalent excess noise $\epsilon_{equ}$, defined from the covariance matrix. 
	
	The eavesdropping model with optical teleportation is shown in Fig. \ref{fig1} (a), where Eve has two attack stations and a shared Einstein-Podolsky-Rosen (EPR) entanglement source $\rho_E$. 
	With this setup, Eve can simulate a equivalent Gaussian channel $\mathcal{G}_{equ}$ with parameters $T_{equ}$ and  $\epsilon_{equ}$, which gives the same covariance matrix as what Alice and Bob have.
	
	The first station (I) is close to Alice, in which Alice’s quantum signal and one mode of EPR source $\rho_E$ with the variance $V_{\rho}$ go through a two-mode squeezing operation $S_g$. Then one output goes through a lossy noisy channel $\mathcal{G}_c$ to Eve’s second station (station II). The channel $\mathcal{G}_c$ can be a normal fiber, or the best low-loss fiber that Eve can access.
	In the second station (II), Eve prepares the second EPR source $\phi_E$ with variance $V_{\phi}$ to help simulating channel noise $\epsilon_{equ}$, by combing with the received mode through a beamsplitter $B_{\eta}$ with transmittance $\eta$. Then one output combines with the received signal from channel $\mathcal{G}_c$ through a second beamsplitter $B_{t}$ with transmittance $t$. One of its output is then sent to Bob. The modes $E_1$, $E_2$ and $E_3$ are stored in the quantum memory and later collectively measured. 
	
	Such a teleportation based attack to simulating a fiber channel $\mathcal{G}_{equ}$ needs to satisfy following constraints, 
	\begin{equation}\label{eq:Tequ}
		T_{equ}=gT_{c}t,
	\end{equation}
	\begin{equation}\label{eq:chi_{equ}}
		\chi_{equ}=t((g-1)T_{c}a+\chi_{c})+(1-t)b-2\sqrt{t(1-t)(g-1)T_c}c.
	\end{equation}
	where $\chi_{equ}$ and $\chi_{c}$ represent the channel output noise with $\chi_{equ}=1-T_{equ}+T_{equ}\epsilon_{equ}$ and $\chi_{c}=1-T_{c}+T_{c}\epsilon_{equ}$ respectively.
	And $a$, $b$, $c$ correspond the components of the covariance matrix $\gamma_{E_{\rho}E_{\phi}}$ describing the modes $E_{\rho}$, $E_{\phi}$ in FIG. \ref{fig1},
	\begin{align}\label{eq:gammaErEp}
		\begin{split}
			&\gamma_{E_{\rho}E_{\phi}}=
			\begin{pmatrix}
				a\cdot I_2 & c\cdot\sigma_z \\
				c\cdot\sigma_z & b\cdot I_2 
			\end{pmatrix}\\=
			&\begin{pmatrix}
				V_{\rho}\cdot I_2 & \eta\sqrt{V_{\rho}^2-1}\cdot\sigma_z \\
				\eta\sqrt{V_{\rho}^2-1}\cdot\sigma_z & \eta{V_{\rho}}+(1-\eta)V_{\phi}\cdot I_2 
			\end{pmatrix}
		\end{split}
	\end{align}
	where $I_2=
	\begin{pmatrix}
		1 & 0 \\
		0 & 1
	\end{pmatrix}$
	,
	$\sigma_z=
	\begin{pmatrix}
		1 & 0 \\
		0 & -1
	\end{pmatrix}$.
	
	In this model, the variation of eavesdropping strength is reflected in the preparation and distribution of Eve's EPR source $\rho_E$ with variance $V_{\rho}(\gamma)=\frac{1+\gamma^2}{1-\gamma^2}$, where $\gamma$ is the squeezing parameter. The minimum variance of $\rho_E$ required to simulate the Gaussian channel $\mathcal{G}_{equ}$ \cite{tserkis2018simulation} is given by
	\begin{equation}\label{eq:Vrhomin}
		V_{\rho}^{min}=\frac{1+\gamma_{min}^2}{1-\gamma_{min}^2},
	\end{equation}
	with 
	\begin{equation}\label{eq:gamma^min}
		\gamma_{min}=\frac{-e-\sqrt{e^2-4df}}{2d}
	\end{equation}
	with the parameters
	\begin{equation}\label{eq:d}
		d=T_{c}t(g-1)+(1-t)+(t\chi_{c}+\chi_{equ}),
	\end{equation}
	\begin{equation}\label{eq:e}
		e=-4\sqrt{t(1-t)(g-1)T_c},
	\end{equation}
	\begin{equation}\label{eq:f}
		f=T_{c}t(g-1)+(1-t)+(t\chi_{c}+\chi_{equ}).
	\end{equation}
	
	Except for the parameter $V_{\rho}$, it is worth mentioning that the ideal model usually takes the two mode squeezing gain $g\rightarrow\infty$, where the shared channel $\mathcal{G}_{equ}$ is completely replaced by the standard teleportation protocol. While the parameters {$\eta,V_{\phi}$} are usually optimized to maximize Eve's ability in the following simulation. Fig. \ref{fig1} (b) illustrates the relationship between the model performance and the variance of Eve's EPR source.
	The simulation shows that the key rate gradually decreases as $V_{\rho}$ increases. When $V_{\rho}$ approaching infinity, the performance of this teleportation-based attack reaches the optimal collective attack. In this case, $B_{\eta}$ has the same transmittance as the channel transmittance $\eta = T_{equ}$, and $V_{\phi}=1+\frac{T_{equ}\epsilon_{equ}}{1-T_{equ}}$.
	When $V_{\rho}$ lies at the minimum $V_{\rho}^{min}$, the model degrades to the individual attack. The transmittance of beamsplitter $B_{\eta}$ is $\eta=1$, meaning no EPR $\phi_E$ is required.
	 
	When $g$ takes a finite value as a more realistic situation, the simulated Gaussian channel $\mathcal{G}_{equ}$ is noisier due to the presence of $\mathcal{G}_{c}$, usually means a weaker eavesdropping strength. The finite $g$ leads to the increase of the required minimum variance $V_{\rho}^{min}$ to simulate channel $\mathcal{G}_{equ}$, which requires better entanglement generation and distribution ability.
	If Eve can prepare the EPR source with infinite entanglement, by taking {$\eta=\frac{T_{equ}-tT_c}{1-t}$}, she can still achieve the optimal collective attack. 
	
	\subsection{Limitation caused by fiber loss and entanglement distillation}
	The above analysis shows that it has requirement for the entanglement of EPR source to achieve the optimal collective attack. However, various imperfections in the entanglement generation and distribution make remotely distribution of high-quality EPR source is a generic difficult problem in practical experiments.
	
	Among all the imperfections, fiber loss matters the most since it will cause a huge entanglement degradation. Fiber loss consists of several parts, mainly including confinement loss, surface scattering loss, Rayleigh scattering, macrobending loss, microbending loss, intra-red absorption loss. Some of these loss are intrinsic which is difficult to be eliminated by current fiber fabrication technologies. In most operators' deployed fiber, the attenuation coefficient is usually counted as 0.275 dB/km high. While standard G.652 fiber is  0.18$\sim$0.2 dB/km, low-loss fiber is of 0.15$\sim$0.17 dB/km. Furthermore, it is believed that in the not-too-distant future a hollow-core antiresonant optical fiber \cite{fokoua2023loss} can further lower the attenuation coefficient to 0.1 dB/km. Even with these best low-loss fibers, the attenuation can still easily reach to 5-10 dB high in 50$\sim$100 km range. 
	
	To overcome the entanglement degradation caused by the fiber loss, an additional entanglement distillation step is needed in station II. For the Gaussian EPR source like the commonly used two-mode squeezed state in CV-QKD, the entanglement distillation usually consists of two steps \cite{campbell2012gaussification, campbell2013continuous}. The first is to improve the entanglement degree with non-Gaussian operations, resulting in non-Gaussian output states. The second is Gaussification, which transform the non-Gaussian state back to a Gaussian state, while keep the entanglement degree still improved. 
		\begin{figure*}[t]
		\includegraphics[width=16cm]{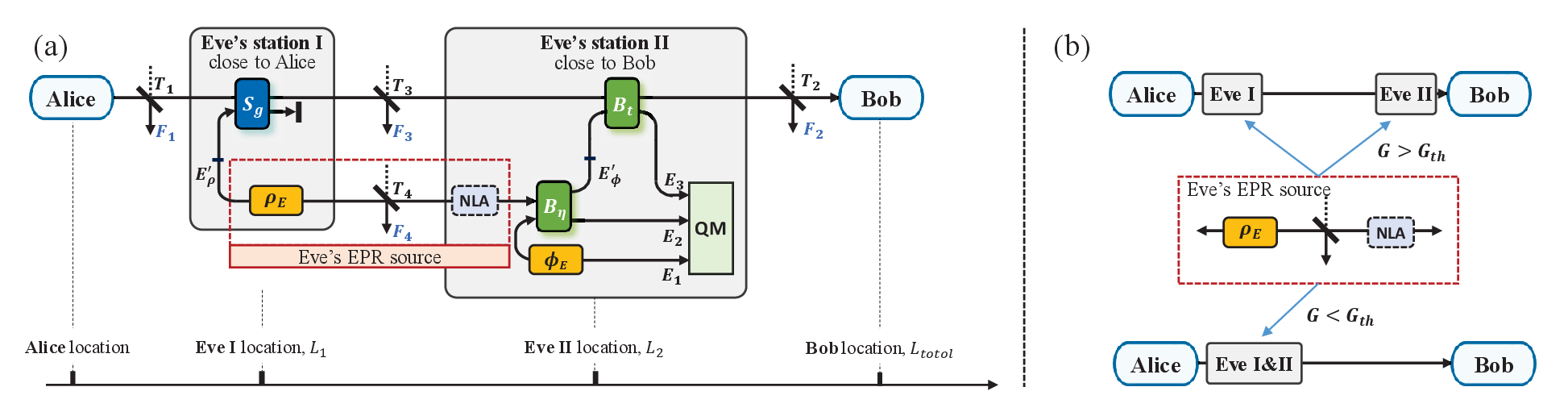}
		\caption{\label{fig2}$(a)$ The EB scheme of the practical teleportation-based eavesdropping model with NLA. The four beamsplitters with transmittances $T_1$, $T_2$, $T_3$, and $T_4$ describe the fiber losses for different fiber links. $L_1$ is the distance from Alice to station I, $L_2$ is the distance from Alice to station II, and $L_{total}$ is the total distance from Alice to Bob. $(b)$ The locations of Eve's stations that enable the optimal eavesdropping. If the gain of NLA is larger enough, $G\textgreater G_{th}$, the optimal choice of station I and II are located close to Alice and Bob, respectively. If the gain of NLA is weak, $G \textless G_{th}$, the optimal choice will be station I and station II are placed together and close to Alice.}
	\end{figure*}

	There is an alternative method to distill the Gaussian EPR, which is probabilistic noiseless linear amplification (NLA) \cite{blandino2012improving, fiuravsek2012gaussian, yang2012continuous}. When amplification succeed, NLA can amplify a coherent state $\ket{\alpha}$ to $\ket{ G\alpha}$ without noise, where $G$ is the gain factor of NLA. When applying NLA to one mode of a two-mode squeezed state, which passed through a lossy channl, the output state can be seen as a new Gaussian state with larger initial entanglement and one mode passing through a channel with less loss. 
		
	In next section, we will analyze the limitation imposed by practical fiber loss on the eavesdropping ability. And we choose NLA as the entanglement distillation method for simplicity. 
	Beside the fiber loss between Eve's two distant stations, the fiber transmission of signal state, including from Alice to station I and from station II to Bob, should be also taken into account.

\section{\label{sec:level3} limited eavesdropping with practical fiber}
	When considering practical fiber loss, the locations of Eve's two stations and how well the entanglement distributed between two stations will influence the eavesdropping strength. The eavesdropping model is described as FIG. \ref{fig2}. Assuming Eve's station I and station II are located at $L_1$ and $L_2$. There are four fiber channels, 1) from Alice to Eve's station I, 2) from Eve's station I to station II for transmitting amplified quantum signal, 3) from Eve's station I to station II for distributing EPR source, 4) from Eve's station II to Bob. All this four channels we assume that Eve can replace them by fibers with better quality, but still having a loss, as discussed above $\alpha = 0.18 \sim 0.2$ dB/km for G.652 fiber, $0.15\sim0.17$ dB/km for low-loss fiber, $ 0.1$ dB/km for hollow-core fiber.  
	
	Eve's distributed EPR is first generated in station I as $\rho_E$, and then one mode is sent to station II through a fiber channel. Then at station II, Eve employs a NLA for entanglement distillation, which means many copies will be consumed to generate one better EPR due to the probabilistic nature of NLA. 
	
	Intuitively, the better the distributed entanglement is, the more information Eve can access. This will put requirements on entanglement distillation, to reduce the entanglement reduction due to fiber loss.  
	
	We will first discuss the situation without NLA, to show the optimal attack strategy when remote entanglement distribution is limited by fiber loss. Then we discuss how NLA improves eavesdropping ability, which eventually can reach optimal collective attack, with two stations separated around Alice and Bob respectively.

\subsection{Eavesdropping without NLA}
In the eavesdropping model without NLA, the locations of Eve's stations are the main consideration. In order to simulate the given channel $\mathcal{G}_{equ}$, the parameters of the model follow these constraints,	
	\begin{equation}\label{eq:T1}
		T_1T_2T_3t{g}=T,
	\end{equation}
	\begin{equation}\label{eq:epsilon1}
		\begin{split}
			&\chi_{equ}=T_2(t(T_{3}(g(1-T_1)+(g-1)a')+(1-T_3))\\
			&+(1-t)b'-2\sqrt{t(1-t)(g-1)T_c}c')+1-T_2.
		\end{split}
	\end{equation}
	
	And $a'$, $b'$, $c'$ correspond the components of the covariance matrix $\gamma_{E_{\rho}^{'}E_{\phi}^{'}}$ describing the modes $E_{\rho}^{'}$, $E_{\phi}^{'}$ in FIG. \ref{fig2},
	\begin{align}\label{eq:gammaErEp'}
		\begin{split}
			&\gamma_{E_{\rho}^{'}E_{\phi}^{'}}=
			\begin{pmatrix}
				a'\cdot I_2 & c'\cdot\sigma_z \\
				c'\cdot\sigma_z & b'\cdot I_2 
			\end{pmatrix}\\=
			&\begin{pmatrix}
				V_{\rho}\cdot I_2 & \sqrt{T_4\eta V_{\rho}^2-1}\cdot\sigma_z \\
				\sqrt{T_4\eta V_{\rho}^2-1}\cdot\sigma_z & \eta{T_4V_{\rho}+(1-T_4)}+(1-\eta)V_{\phi}\cdot I_2 
			\end{pmatrix}
			.
		\end{split}
	\end{align}
	
	To focus on the influence of locations of Eve's stations, we set the gain $g$ of two-mode squeezing and the variance of prepared EPR state $\rho_E$ to be sufficiently large in the following simulations. And we assume Eve exploits the hollow-core fiber with attenuation coefficient 0.1 dB/km.
	
\begin{figure}[htbp]
	\includegraphics[width=7cm]{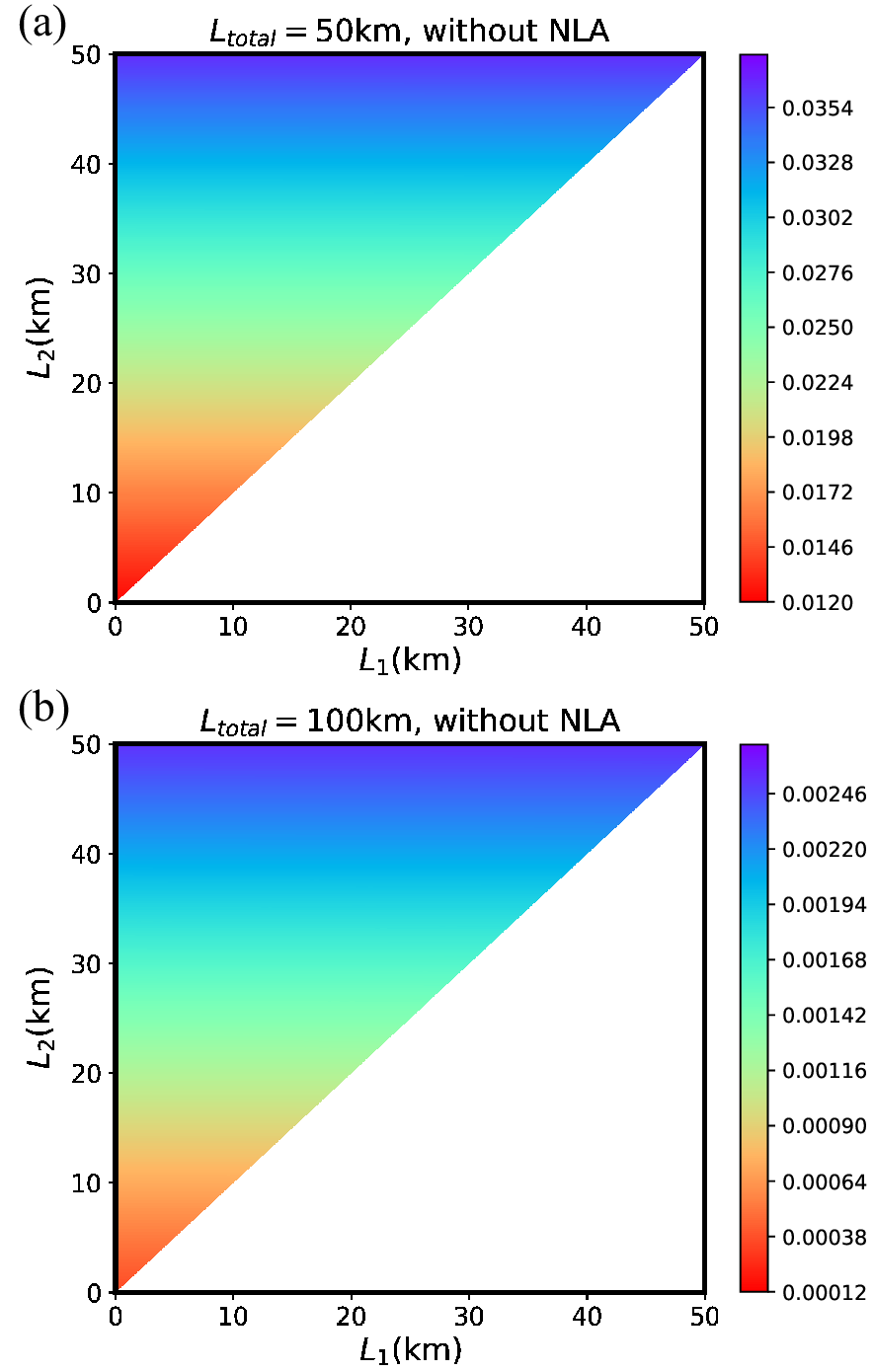}
	\caption{\label{L2-L1_withoutNLA}The influence of Eve's two stations' locations on the secret key rate, when the total distance is 50 km and 100 km, respectively. The simulated channel $\mathcal{G}_{equ}$ has attenuation coefficient 0.275 dB/km, and the excess noise is $\epsilon=0.04$. Other simulation parameters are Alice's modulation variance $V_A=4$ and reconciliation efficiency $\beta=0.96$.}
\end{figure}
\begin{figure}[htbp]
	\includegraphics[width=7cm]{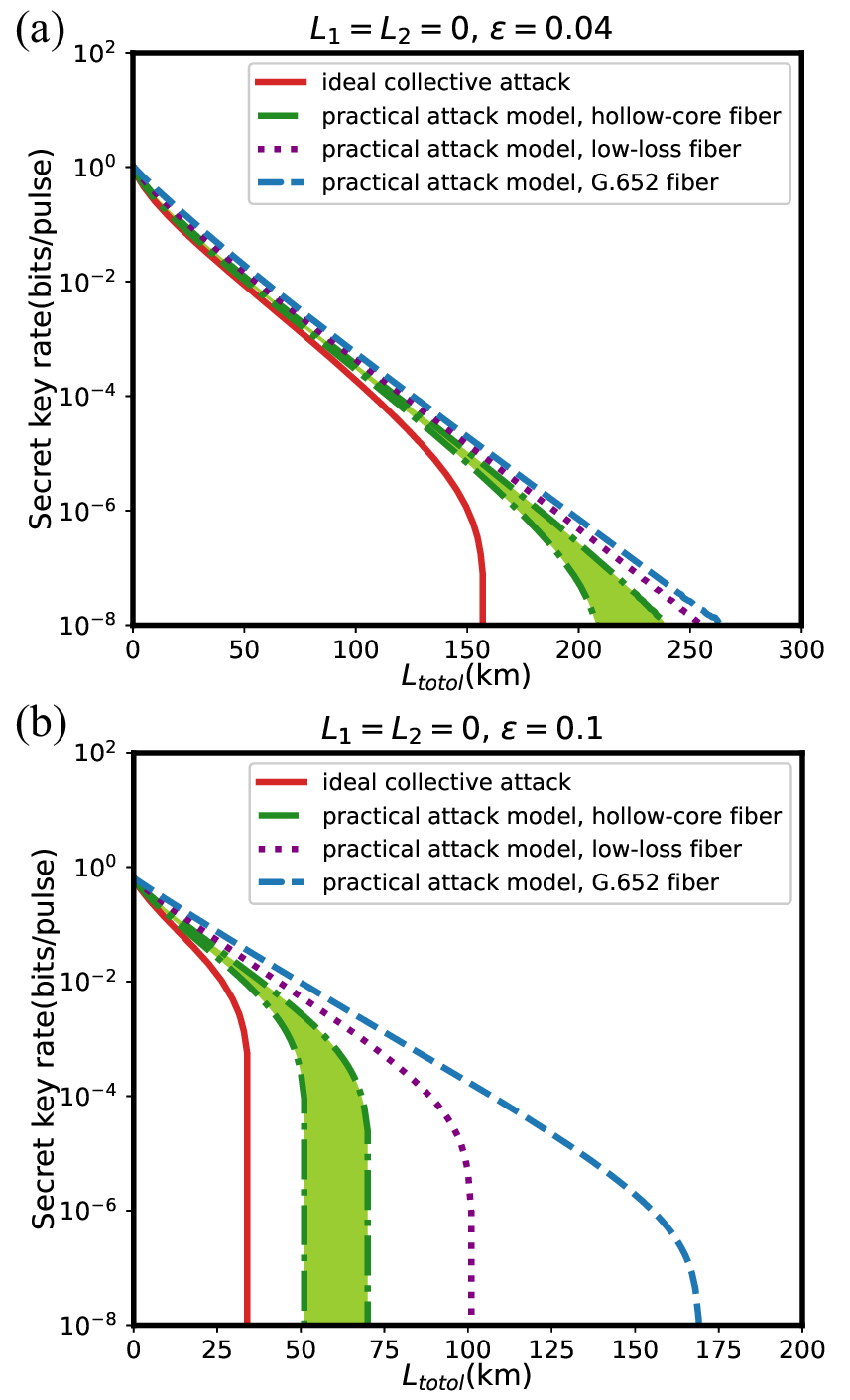}
	\caption{\label{SKR_withoutNLA}Secret key rate vs. transmission distance with different fibers when the system excess noise is 0.04 and 0.1, respectively. The red solid line represents the optimal collective attack case, where Eve can replace the channel with a lossless fiber. The blue dashed line represents the G.652 fiber with 0.2 dB/km, the purple dotted line represents the low-loss fiber with 0.15 dB/km, and the green dot-dashed line represents the best fiber predicted with current theory as attenuation coefficient is around 0.05$\sim$0.1 dB/km. Other simulation parameters are Alice's modulation variance $V_A=4$ and reconciliation efficiency $\beta=0.96$.}
\end{figure}
Firstly, we change both the locations of Eve's two stations $L_1$ and $L_2$ ($L_1\leq L_2\leq{L_{total}}$), to see how they influence the secret key rate. In Fig. \ref{L2-L1_withoutNLA} $(a)$ and $(b)$, we fix total distance to 50 km and 100 km, respectively. It is shown that, when the location $L_1$ of station I is fixed, secret key rate decreases as $L_2$ moves closer to $L_1$, which indicates that $L_2 = L_1$ gives the worst case. In this case, there is actually only one eavesdropping station, and it reduces to an entangling-cloning attack model with a reduced wire-tapping ratio. Besides, when $L_2 = L_1$, the secret key rate decreases as $L_1$ moves closer to Alice, which shows the overall worst case is both Eve's stations are located close to Alice, $L_2 = L_1 = 0$. This follows the intuition that wiretapping at location closer to transmitter site is better, since more signal energy can be harvested. The simulation parameters for Fig. \ref{L2-L1_withoutNLA} $(a)$ and $(b)$ can be found in the figure caption.       

Secondly, we show the secret key rates when Eve chooses $L_2 = L_1 = 0$ and substitutes the rest fiber with different kinds of fibers, as in Fig. \ref{SKR_withoutNLA} $(a)$ and $(b)$. Comparing to the ideal collective attack (red line), Eve's practical eavesdropping ability is limited by the intrinsic fiber loss. The better fiber that Eve can use, the stronger eavesdropping ability she has. For a common QKD system with $\epsilon=0.04$, under the $50$ km transmission distance, the secret key rate could be increased by $20\%-40\%$ when Eve uses the best-expected hollow-core fiber. When Eve has no choice but the commonly used G.652 fiber, secret key rate could be increased even by $140\%$. It can be seen that if considering a practical eavesdropper with limited fiber technology, there will be a significant improvement in the secret key rate.
The increase in transmission distance is more remarkable in the noisier system with $\epsilon=0.1$.
In $(b)$, the transmission distance can be extended from the original 30 km to a maximum of 170 km and a minimum of 50 km, which is very meaningful for the practical application of QKD systems.
	\begin{figure*}[htbp]
	\includegraphics[width=16cm]{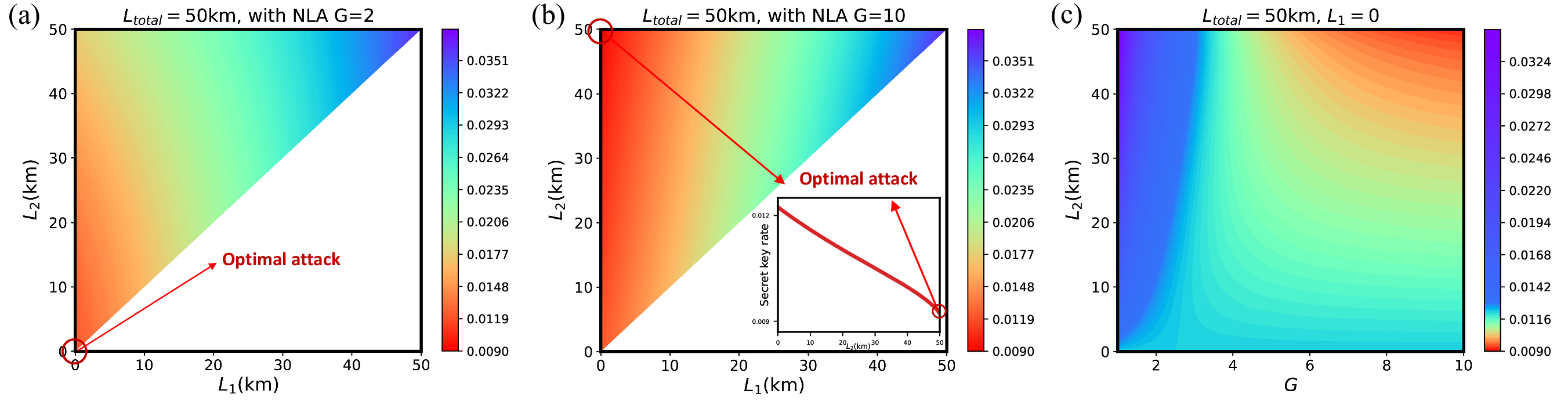}
	\caption{\label{L2-L1_NLA}$(a)$ and $(b)$, the influence of Eve's two stations' locations on the secret key rate, when the gain of NLA are $G=2$ and $G=10$, respectively. Assume Eve uses the hollow-core fiber with attenuation coefficient of 0.1 dB/km. $(c)$ Secret key rate when change the location of station II $L_2$ and the gain $G$ of NLA, when station I is placed at the transmitter ($L_1=0$). For each G, the initial EPR state is optimized to make the equivalent source $\rho_E^{G}$ with infinite entanglement $(\gamma^G \rightarrow 1)$. Other parameters remain the same as before.}
\end{figure*}

In the above analysis, even Eve can prepare ideal EPR source with infinite entanglement, its practical ability is still limited due to the entanglement reduction caused by the transmission loss, when distributing one mode of the EPR to the station II. And this is actually the reason leads to the conclusion that both stations should locate at the same site to achieve the optimal attack, under which the system performance still shows a great improvement compared with the ideal model.
At the same time, Eve is also able to take many measures to resist the influence introduced by the fiber loss.
One natural method to improve the eavesdropping ability is having entanglement distillation in station II, this will be discussed in next. 
\subsection{Eavesdropping with NLA}
We consider using NLA as entanglement distillation method, which simplifies the analysis for Gaussian channel while keeping the main conclusion. 

NLA is a probabilistic operation, which can amplify a coherent state $\ket{\alpha}$ to $\ket{G \alpha}$ without noise when amplification succeed, $G$ is the gain factor of NLA. When applying NLA to one mode of a two-mode squeezed state $\rho_E$ with squeezing parameter $\gamma$, which passed through a lossy channel with transmittance $T_4$, the output state's entanglement will be improved. When considering the output is still a Gaussian state for convenience, it is equivalent to a new two-mode squeezed state with sqeezing parameter $\gamma^G$ passing through a lossy channel with transmittance $T_4^G$, which fulfill following conditions with
\begin{equation}\label{eq:NLAgamma}
	\gamma^{G}=\gamma\sqrt{1+(G^2-1)T_4},
\end{equation}
\begin{equation}\label{eq:NLAT4}
	T_4^{G}=\frac{G^2T_4}{1+(G^2-1)T_4}.
\end{equation}
\begin{figure*}[htbp]
	\includegraphics[width=16cm]{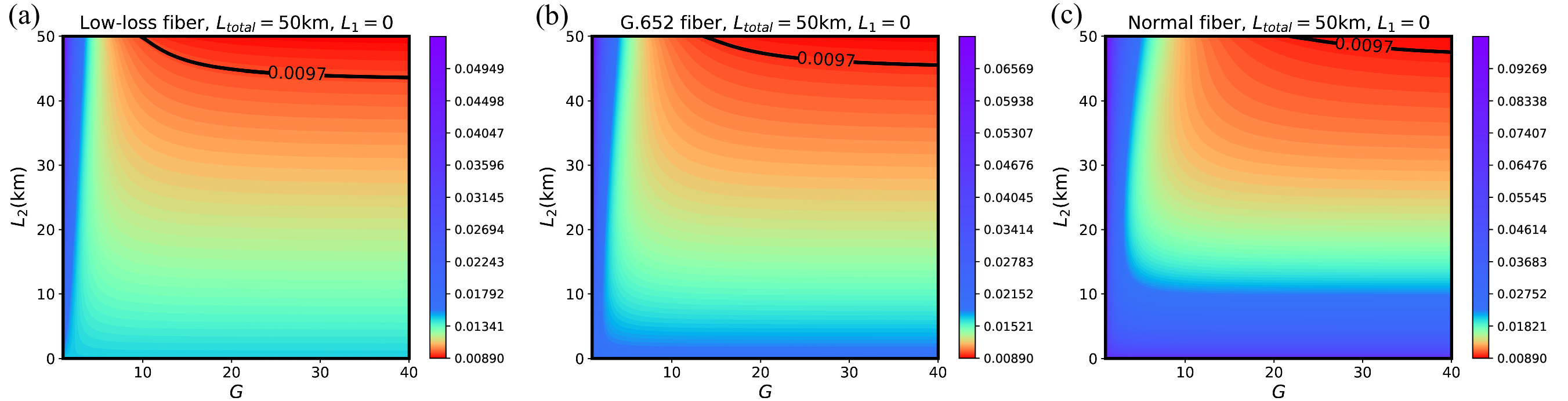}
	\caption{\label{L2-G_fiberloss} Secret key rates when Eve uses different types of fibers, where the total distance is 50 km. $(a)$ Low-loss fiber with 0.15 dB/km, $(b)$ G.652 fiber with 0.2 dB/km and $(c)$ normal fiber with 0.275 dB/km in network.
		The upper right corner at each figure represent the case that the secret key rate is closing the worst-case, which is the optimal collective attack. The simulation parameters remain the same as before.}
\end{figure*}
This gives an upper bound on the amplification gain, when keeping the output still Gaussian form. Besides, in order to simulate the given channel $G_{equ}$, the equivalent squeezing parameter $\gamma^{G}$ also has a restriction as in Section II,
	
	\begin{equation}\label{eq:NLAres1}
		\gamma^{G}_{min}\leq\gamma^{G}\leq 1.
	\end{equation}
	Here the minimum squeezing parameter $\gamma^{G}_{min}$ is re-defined as
	\begin{equation}\label{eq:NLAres2}
		\gamma_{min}^{G}=\frac{-e_{G}-\sqrt{{e_{G}}^2-4d_{G}f_{G}}}{2d_{G}}
	\end{equation}	
with the parameters
	\begin{equation}\label{eq:d}
		d_{G}=\frac{T_{equ}}{T_2}+2T_4^G-1+\frac{\chi_{equ}-(1-T_2)}{T_2},
	\end{equation}
	\begin{equation}\label{eq:e}
		e_{G}=-4\sqrt{\frac{T_{equ}}{T_1T_2}T_4^G},
	\end{equation}
	\begin{equation}\label{eq:f}
		f_{G}=2\frac{T_{equ}}{T_1T_2}-\frac{T_{equ}}{T_2}+1-\frac{\chi_{equ}-(1-T_2)}{T_2}.
	\end{equation}
	
We consider two cases for the rest simulations, I) optimizing the initial EPR state $\rho_E$ to make the equivalent source $\rho_E^{G}$ with infinite entanglement $(\gamma^G \rightarrow 1)$ to see the optimal distilled case, and II) fixing the initial entanglement of $\rho_E$ to see how does the gain of NLA influence.
	
	First, we consider the simulation of case I. Optimizing initial EPR to make the distilled $\gamma_G \rightarrow 1$ actually means, for each NLA gain G, the distilled output state keeps a same EPR source while a different channel loss $T_4^G$. From Eq.(\ref{eq:NLAT4}), larger G means higher $T_4^G$, therefore better distributed EPR entanglement. 
	
	In Fig. \ref{L2-L1_NLA} $(a)$ and $(b)$, we move both locations of Eve's two stations simultaneously, with $G=2$ in $(a)$ and $G=20$ in $(b)$, while keeping the total distance $L_{total}=50$ km unchanged. Both $(a)$ and (b) show that, the strongest eavesdropping happens when station I is located at Alice's end ($L_1=0$), while the situation is different regarding to the station II's location. As $(a)$ has a weaker NLA gain, the worst-case secret key rate happens when Eve have both station I and II placed on the transmitter side, which is consistent with the previous discussion without NLA. While $(b)$ has a larger NLA gain, one can find that, on opposite as $(b)$, the optimal eavesdropping would preferably have station I and II  placed on the transmitter and receiver side, respectively.
	
	This conjecture is further verified in FIG. \ref{L2-L1_NLA} $(c)$, where station I is placed at the sending side directly and only station II is moved. It shows that, for each NLA gain G, the secret key rate varies with different location of station II. And there exists a threshold $G_{th}$, when $G<G_{th}$, the optimal eavesdropping happens at the condition both Eve's stations locate at transmitter side ($L_2=L_1=0$), while $G>G_{th}$, the optimal eavesdropping happens at the condition Eve's two stations separately locate at transmitter side (station I, $L_1 = 0$) and receiver side (station II, $L_2 = L_{total}$). This is schematically illustrated in FIG. \ref{fig2} $(b)$.
	
	The above simulation is based on hollow-core fiber ($\alpha = 0.1$ dB/km). We also investigate different fibers with higher loss that Eve could use, which show the same feature as hollow-core fiber, as shown in FIG. \ref{L2-G_fiberloss}. No matter how worse the fiber loss is, as long as the gain of NLA is large enough, the optimal eavesdropping strategy is the same, i.e., Eve's two stations separately locate at transmitter side and receiver side. And when G is large enough, the eavesdropping ability is upgraded closing to the optimal collective eavesdropping. 
	 
	\begin{figure}[htbp]
		\includegraphics[width=7cm]{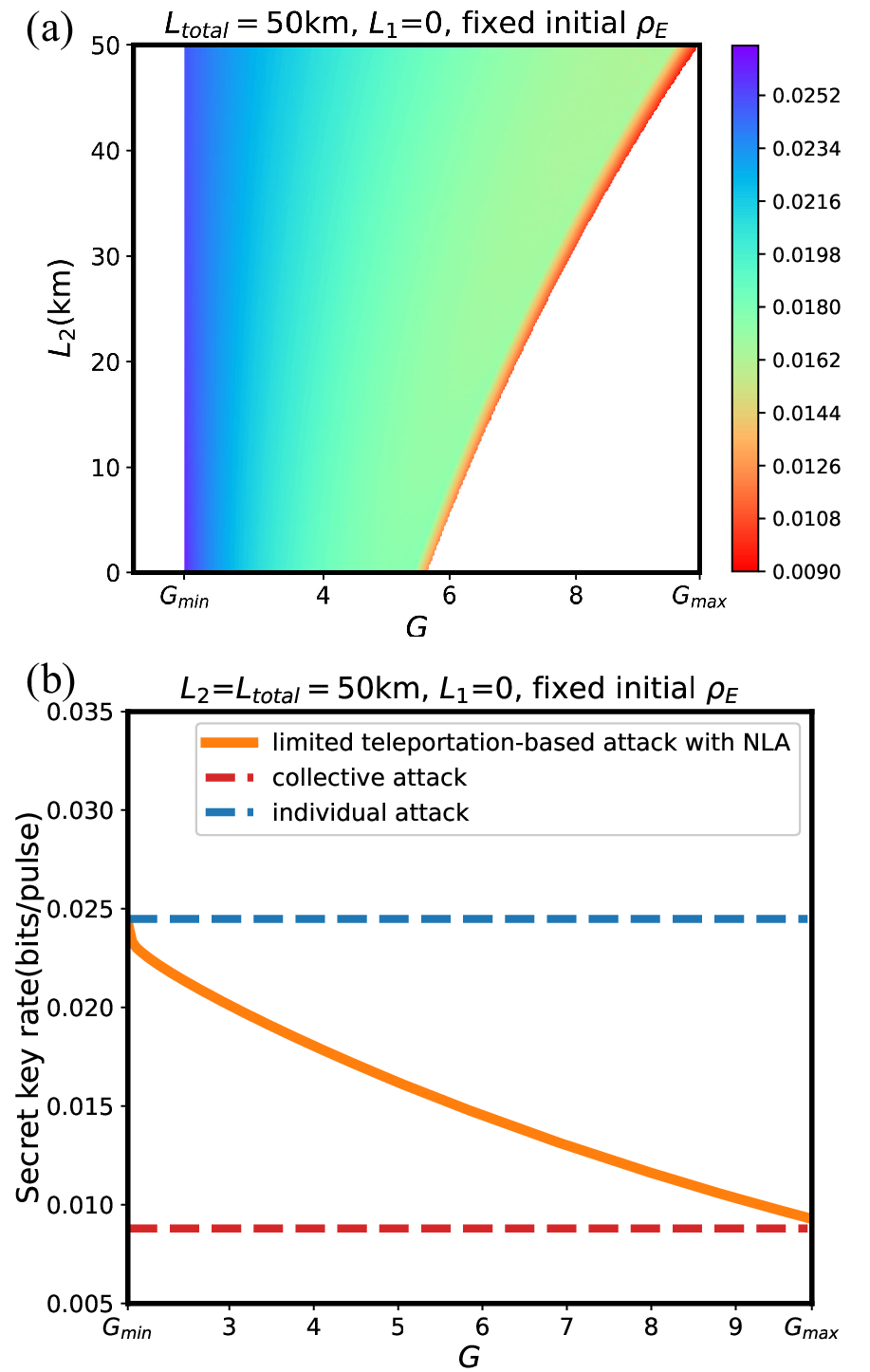}
		\caption{\label{fixed-variance}(a) The influence of station II's location and the gain $G$ on the secret key rate, when the variance $V_{\rho}$ of initial EPR state $\rho_E$ is fixed. (b) The secret key rate vs. NLA gain $G$, when $V_{\rho}$ is fixed. Other parameters remain the same as before.}
	\end{figure}
	
	We next investigate the simulation case II with fixed initial EPR state $\rho_E$, which is more technologically reasonable since current two-mode squeezed state generation is not mature enough to easily generate arbitrary large entanglement. We still assume station I is located at the transmitter side. For each station II location $L_2$, the NLA gain will be limited by the condition $\gamma^G_{min}\le\gamma^G\le 1$, derived from Eq.(\ref{eq:NLAgamma}) and (\ref{eq:NLAres1}).
	The simulation results are in Fig. \ref{fixed-variance}, the left boundary in $(a)$ lies on the condition $\gamma^G=\gamma^G_{min}$ indicating the individual attack, and the right boundary lies on the condition $\gamma^G=1$,  indicating different kinds of collective attack.
	The secret key rate of the left boundary is consistent with the optimal individual attack, while the key rate of the right boundary decreases with the increase of $L_2$. When $L_2=L_{total}$, the eavesdropping at this point is close to the optimal collective attack.
	
	This simulation shows that the location of eavesdropping will have an effect on the strength of the collective attack while the individual attack is not influential.
	Fig. \ref{fixed-variance} $(b)$ clearly demonstrates the compensatory effect of the NLA for limited eavesdropping performance.
	Picking a appropriate EPR state, the entanglement of Eve's distributed EPR source can still be improved so that the limited eavesdropping with lossy fiber can be converted from the individual attack to the optimal collective attack, with specific NLA. 
	
	From the above simulations, it is clear that the distillation of entanglement source is the crucial factor that affects the eavesdropping ability.
	The presence of practical fiber loss will limit the distributed entanglement, resulting in the optimal attack occurring when Eve's two stations are merged and placed at the transmitter.
	The practical QKD system performance under such the optimal attack is better as the fiber loss is higher.
	When NLA is used to greatly optimize the distillation of the entanglement source, the optimal attack will occur when the two stations are separated on both the transmitting and receiving sides, where it can even approach the optimal collective attack when the NLA gain is large.
	\section{Discussion}
	All-optical-teleportation-based attack model is of interest for allowing that Eve doesn't have to fully control the shared channel, in which the entanglement source is the key-role to eavesdropping ability.
	In the ideal eavesdropping model, Eve is usually assumed to have an infinite power to prepare, distill, and distribute arbitrary entanglement sources.
	The gap between realistic and ideal scenarios will challenge the premise, among which the most important and inevitable imperfection is the practical fiber loss. Indeed, Eve can only replace the normal channel between the communicating parties with a low-loss channel as far as possible, such as hollow-core fiber.
	We have found that in the realistic environment, even if Eve can use the best available hollow-core fiber to perform eavesdropping, the key rate and transmission distance of QKD system will still be improved to a greater extent.
	In addition to the amount of fiber loss, the location of fiber loss also has an impact on Eve's eavesdropping ability. 
	
	We also analyzed when NLA is used for entanglement distillation to improve the practical eavesdropping ability.
	Numerical simulations reveal that fiber loss will challenge Eve's ability, while the NLA can be used to improve her eavesdropping strength. After the compensation from NLA on the distributed entanglement, the eavesdropping ability could even be improved approaching the performance of optimal collective attack.
	
	Despite of this, we believe that other non-ideal factors during entanglement distribution, such as finite entanglement generation, imperfect squeezing with limited gain, etc., should also be investigated in the future, further deeper the understanding of realistic risk of practical CV-QKD system.
	
	\begin{acknowledgments}
	This research was supported by the National Natural Science Foundation of China (62001044 and 61531003), the Basic Research Program of China (JCKY2021210B059), the Equipment Advance Research Field Foundation (315067206), and the Fund of State Key Laboratory of Information Photonics and Optical Communications (IPOC2021ZT02).
	\end{acknowledgments}
	\appendix
	\section{\label{appendixA}The Entanglement of Eve's source.}
	In the entangling cloning eavesdropping model for the Gaussian CV-QKD, Alice first prepares an EPR state $\hat{\Psi}_{AA'}$ whose covariance matrix has the form
	\begin{align}\label{eq:Psi}
		\gamma_{AA'}=
		\begin{pmatrix}
			V\cdot I_2 & \sqrt{V^2-1}\cdot\sigma_z \\
			\sqrt{V^2-1}\cdot\sigma_z & V\cdot I_2 
		\end{pmatrix}
		,
	\end{align}
	with $V=V_A+1$. 
	One mode of the EPR state $\hat{\Psi}_{AA'}$ $A'$ is transmitted through a given Gaussian channel $\mathcal{G}_{equ}(T_{equ},\epsilon_{equ})$ as $B$ to Bob, where its covariance matrix transform of is described as, 
	\begin{align}\label{eq:PsiAB}
		\Psi_{AB}=\mathcal{G}(\Psi_{AA'})=\mathcal{T}\Psi_{AA'}(\mathcal{T})^T+\mathcal{N}.
	\end{align}
	$\mathcal{T}$ and $\mathcal{N}$ characterize the transmittance and the excess noise, respectively. The covaroance matrix $\gamma_{AB}$ is given by,
	\begin{align}\label{eq:gammaAb}
		\gamma_{AB}=
		\begin{pmatrix}
			V\cdot I_2 & \sqrt{T_{equ}(V^2-1)}\cdot\sigma_z \\
			\sqrt{T_{equ}(V^2-1)}\cdot\sigma_z & (T_{equ}V+(1-T_{equ})N)\cdot I_2 
		\end{pmatrix}
	\end{align}
	where $N=1+\frac{T_{equ}\epsilon_{equ}}{1-T_{equ}}$ represents the noise variance.
	
	In the all-optical teleportation model, the signal mode $A'$ will pass through $S_g$, Gaussian channel $\mathcal{G}_c$ and $B_t$ successively with symplectic transformations as
	\begin{align}\label{eq:Sg}
		S_g=
		\begin{pmatrix}
			\sqrt{g}\cdot I_2 & \sqrt{g^2-1}\cdot I_2 \\
			-\sqrt{g^2-1}\cdot I_2 & \sqrt{g}\cdot I_2 
		\end{pmatrix}
		,
	\end{align}
	\begin{align}\label{eq:Bt}
		B_t=
		\begin{pmatrix}
			\sqrt{t}\cdot I_2 & \sqrt{1-t}\cdot \sigma_z \\
			\sqrt{1-t}\cdot\sigma_z & \sqrt{t}\cdot I_2 
		\end{pmatrix}
		.
	\end{align}
	The covariance matrix of its output $B'$ and mode $A$ $\gamma_{AB'}$ is given as Eq.\ref{eq:gammaAb'}.
	\begin{figure*}[t]
		\includegraphics[width=14cm]{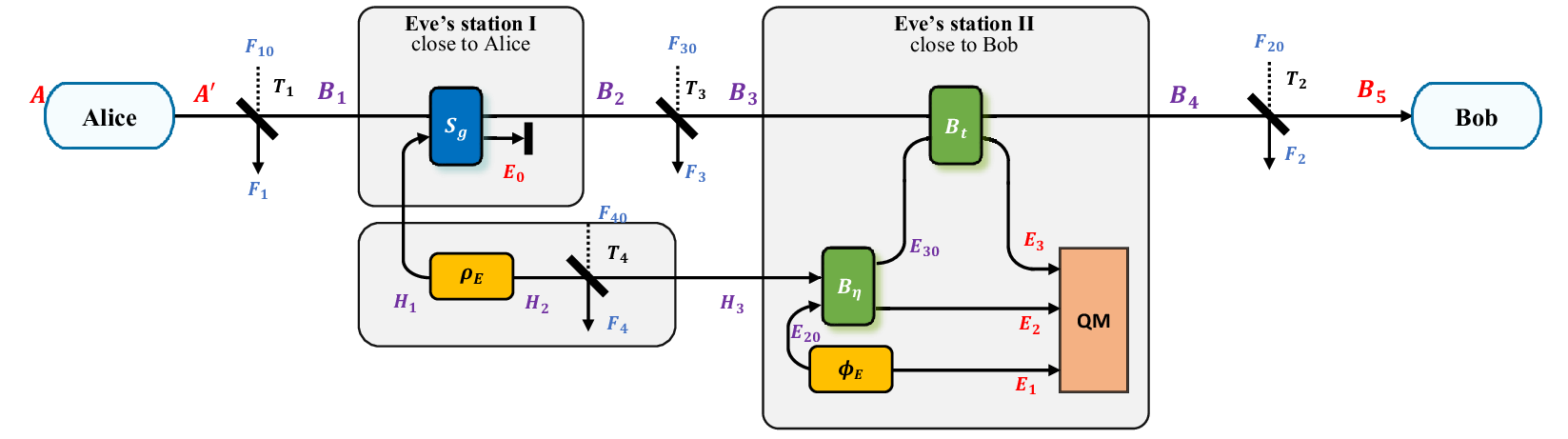}
		\caption{\label{APPB} The mode transformation of the all-optical teleportation with practical fiber loss.
		The mode $A$ is reserved by Alice, while $B_5$ is the pattern received by Bob.
		Eve performs the collective attack on $E_1$, $E_2$ and $E_3$ in quantum memory to steal secret keys.
		It should be stressed that since the channel loss is not under the control of Eve, the channel modes $F_1$, $F_2$, $F_3$ and $F_4$ modelled by beamsplitters with transmittances $T_1$, $T_2$, $T_3$ and $T_4$ are trusted, resulting in the increase of the secret key rate.}
	\end{figure*}
	\begin{figure*}[t]
		\begin{equation}\label{eq:gammaAb'}
			\gamma_{AB'}=
			\begin{pmatrix}
				V\cdot I_2 & \sqrt{tT_cg(V^2-1)}\cdot\sigma_z \\
				\sqrt{tT_cg(V^2-1)}\cdot\sigma_z & (T_{equ}V+t((g-1)T_{c}a+\chi_{c})+(1-t)b-2\sqrt{t(1-t)(g-1)T_c}c)\cdot I_2 
			\end{pmatrix}
			.
		\end{equation}
	\end{figure*}In order to adequately describe the given Gaussian channel, the output of the all-optical teleportation model should be the same as the previous model, that is, Eq.\ref{eq:gammaAb'} should be equivalent to Eq.\ref{eq:gammaAb}, so there is
	\begin{equation}\label{eq:Tequ_A}
		T_{equ}=gT_{c}t,
	\end{equation}
	\begin{equation}\label{eq:chi_{equ}_A}
		\chi_{equ}=t((g-1)T_{c}a+\chi_{c})+(1-t)b-2\sqrt{t(1-t)(g-1)T_c}c.
	\end{equation}

	The EPR source $\rho_E$ in the all-optical teleportation model has a minimum entanglement as $\eta=1$ where the channel noise is completely describes by the EPR state source as the individual attack, then the the covariance matrix $\gamma_{E_{\rho}E_{\phi}}$ describing the modes $E_{\rho}$, $E_{\phi}$ is converted to
	\begin{align}\label{eq:gammaErEp}
		\begin{split}
			&\gamma_{E_{\rho}E_{\phi}}\\=
			&\begin{pmatrix}
				V_{\rho}^{min}\cdot I_2 & \sqrt{{V_{\rho}^{min}}^2-1}\cdot\sigma_z \\
				\sqrt{{V_{\rho}^{min}}^2-1}\cdot\sigma_z & \eta{V_{\rho}}\cdot I_2 
			\end{pmatrix}
			.
		\end{split}
	\end{align}
	$V_{\rho}^{min}$ can be solved naturally.
	Correspondingly, when the EPR source has infinite entanglement where its variance is infinite, Eq.\ref{eq:chi_{equ}_A} holds when $\eta=T_{equ}$ and $V_{\phi}=N$.
	At this point, the channel excess noise is fully characterized by Eve's station II as the collective attack.
	\section{\label{appendixB}The Secret Key Rate of Limited Eavesdropping with Fiber Losses.}
	The secret key rate formula against the collective attack is given by the Devetak-Winter formula,
	\begin{equation}\label{eq:keyrate0}
		R=\beta I{(a:b)}-{S}{(x:E)},
	\end{equation}
	where $\beta$ is the reconciliation efficiency \cite{van2004reconciliation}.
	$I(a:b)$ is the mutual information between Alice and Bob, and ${S}(x:E)$ is the mutual information between Alice and Eve. The amount of information Eve can extract ${S}{(x:E)}\leq{\chi}{(x:E)}$ where ${\chi}{(x:E)}$ represents the Holevo bound.
	The maximum amount of secret keys for Eve to perform the eavesdropping  can be given by
	\begin{equation}\label{eq:Holevo}
		{S}{(x:E)}=S(E)-S(E|x),
	\end{equation}
	which can be obtained through the symplectic eigenvalues of a N-mode state.
	
	In the practical environments, the pure fiber losses are all modeled as a beamsplitter, such as
	\begin{align}\label{eq:BeT}
		B_{T_1}=
		\begin{pmatrix}
			\sqrt{T_1}\cdot I_2 & \sqrt{1-T_1}\cdot \sigma_z \\
			\sqrt{1-T_1}\cdot\sigma_z & \sqrt{T_1}\cdot I_2
		\end{pmatrix}
		.
	\end{align}
	The parameters $T_2$, $T_3$, $T_4$ in the model are also described in this way.
	
	The EB model with mode transformation is shown in FIG. \ref{APPB}. 
	The output after the first channel loss is represented as $B_1$ where $F_{10}$ is the vacuum state,
	\begin{equation}\label{eq:B1}
		\gamma_{AB_1F_1}=B_{T_1}^{A'F_{10}}(\gamma_{AA'}\oplus I_2)(B_{T_1}^{A'F_{10}})^T
		,
	\end{equation}
	here $B_{T_1}^{A'B_1}=I_2\oplus B_{T_1}$.
	After the two mode squeezing operation $S_g$,
	\begin{equation}\label{eq:B2}
		\gamma_{AB_2E_0H_{3}}=S_{g}^{B_1H_1}(\gamma_{AB_1}\oplus \gamma_{H_1H_{3}})(S_{g}^{B_1H_1})^T
		,
	\end{equation}
	
	\begin{align}\label{eq:H1E30}
		\begin{split}
			&\gamma_{H_1H_{3}}=\\
			&\begin{pmatrix}
				V_{\rho}^{min}\cdot I_2 & \sqrt{T_4({V_{\rho}^{min}}^2-1)}\cdot\sigma_z \\
				\sqrt{T_4({V_{\rho}^{min}}^2-1)}\cdot\sigma_z & (T_4V_{\rho}+1-T_4)\cdot I_2 
			\end{pmatrix}
			,
		\end{split}
	\end{align}
	and $S_{g}^{B_1H_1}=I_2\oplus S_{g}\oplus I_2$.
	Further, the signal will undergo the pure loss with transmittance $T_3$,
	\begin{equation}\label{eq:B3}
		\gamma_{AH_3B_3F_3}=B_{T_3}^{B_2F_{30}}(\gamma_{AH_3B_2}\oplus I_2)(B_{T_3}^{B_2F_{30}})^T
	\end{equation}
	where $B_{T_3}^{B_2F_{30}}=I_2\oplus I_2\oplus B_{T_3}$.
	On the other hand, the mode $H_3$ also passes through the $B_\eta$,
	\begin{equation}\label{eq:E30}
		\gamma_{AB_3E_{30}E_2E_1}=B_{\eta}^{H_3E_{20}}(\gamma_{AB_3H_3}\oplus \gamma{E_{20}}E_1)(B_{\eta}^{H_3E_{20}})^T
	\end{equation}
	where $B_{\eta}^{H_3E_{20}}=I_2\oplus I_2\oplus B_{\eta}\oplus I_2$.
	Here
	\begin{align}\label{eq:Beta}
		B_{\eta}=
		\begin{pmatrix}
			\sqrt{\eta}\cdot I_2 & \sqrt{1-\eta}\cdot \sigma_z \\
			\sqrt{1-\eta}\cdot\sigma_z & \sqrt{\eta}\cdot I_2
		\end{pmatrix}
	\end{align}
	and
	\begin{align}\label{eq:Phi}
		\gamma_{E_{20}E_1}=
		\begin{pmatrix}
			V_{\phi}\cdot I_2 & \sqrt{V_{\phi}^2-1}\cdot\sigma_z \\
			\sqrt{V_{\phi}^2-1}\cdot\sigma_z & V_{\phi}\cdot I_2 
		\end{pmatrix}
		.
	\end{align}
	The modes $B_3$ and $E_{30}$ will together gh through $B_t$,
	\begin{equation}\label{eq:B4}
		\gamma_{AB_4E_3E_2E_1}=B_{t}^{B_3E_{30}}(\gamma_{AE_2E_1B_3E_{30}})(B_{t}^{B_3E_{30}})^T
	\end{equation}
	where $B_{t}^{B_3E_{30}}=I_2\oplus I_2\oplus I_2\oplus B_{t}$.
	At last, there exist the fiber loss between the station II and Bob,
	\begin{equation}\label{eq:B5}
		\gamma_{AE_3E_2E_1B_5F_2}=B_{T_2}^{B_4F_{20}}(\gamma_{AE_3E_2E_1B_4}\oplus I_2)(B_{T_2}^{B_4F_{20}})^T
	\end{equation}
	where $B_{T_2}^{B_4F_{20}}=I_2\oplus I_2\oplus I_2\oplus I_2\oplus B_{T_2}$.
	The modes $E_1$, $E_2$ and $E_3$ are stored in quantum memory controlled by Eve, the secret key rate further can be calculated from Eq.\ref{eq:B5}.
	When $T_1, T_2, T_3, T_4 = 1$, $\gamma_{AB_5}$ in Eq.\ref{eq:B5} degenerates to Eq.\ref{eq:gammaAb'}, at which point the practical model also reduces to the ideal all-optical teleportation-based model.

	\nocite{*}
	
	\bibliography{reference}
	
\end{document}